\input{aipcheck}

\documentclass[
    ,final            
  ]
  {aipproc}

\layoutstyle{8x11double}
\usepackage{subfigure}
\usepackage{epsfig}

\begin{document}

\title{Evidence of increasing acoustic emissivity at high frequency with solar cycle 23 in Sun-as-a-star observations}


\author{R.Simoniello}{
  address={PMOD/WRC Switzerland Physikalisch-Meteorologisches Observatorium Davos-World Radiation Center, 7260 Davos Dorf, Switzerland}
}

\author{W.Finsterle}{
  address={PMOD/WRC Switzerland Physikalisch-Meteorologisches Observatorium Davos-World Radiation Center, 7260 Davos Dorf, Switzerland}
}

\author{R.A.Garc\'ia}{
  address={Laboratoire AIM, CEA/DSM-CNRS-Universit\'e Paris Diderot; CEA, IRFU, SAp, centre de Saclay, F-91191, Gif-sur-Yvette, France}}
  
\author{D.Salabert}{
  address={ IAC, Instituto de Astrofis\'\i ca de Canarias, 38205, La Laguna, Tenerife, Spain}}

\author{A.Jim\'enez}{
  address={IAC, Instituto de Astrofis\'\i ca de Canarias, 38205, La Laguna, Tenerife, Spain }}
\begin{abstract}
 We used long high-quality unresolved (Sun-as-a-star observations) data collected by GOLF and VIRGO instruments on board the ESA/NASA SOHO satellite  to investigate the amplitude variation with solar cycle 23 in the high-frequency band (5.7$<\nu<$6.3 mHz). We found an enhancement of acoustic emissivity over the ascending phase of about 18$\pm$3 in velocity observations and a slight enhancement of 3$\pm$2 in intensity. Mode conversion from fast acoustic to fast magneto-acoustic waves could explain the enhancement in velocity observations. These findings open up the possibility to apply the same technique to stellar intensity data, in order to investigate stellar-magnetic activity.
\end{abstract}

\maketitle

\section{Introduction}
{\bf Hi}gh-frequency {\bf P}eak{\bf S} (HiPS) well above the acoustic cut off frequency ($\nu_{ac}$) of the solar $p$-modes have already been observed in high-degree observations of the Sun. A peak structure up to 9mHz was found, well beyond $\nu_{ac}$ \cite{duv91,jef,lib88}. Furthermore they have been also observed in integrated-sunlight measurements, despite their traveling nature \cite{cha03,gar98,jim05}.
 They have  been named HiPS (High-frequency Peaks) and they consists of equally spaced peaks of almost 140mHz. This finding opens up a unique chance to investigate the amplitude variation of the HiPS over solar cycle 23 in integrated sunlight measurements. 
\section{High frequency Peaks (HiPS)}
\subsection{Intensity and velocity observations}
Fig.1 shows the presence of peaks in the frequency range of 5.5 up to 8mHz as seen from continuum intensity observations provided by the $\bf V$ariability $\bf IR$radiance $\bf G$lobal $\bf O$scillation photometer (VIRGO) \cite{fro1}. It comprises three Sun photometers used independently at 402nm(blue), 500nm(green) and 862nm(red). The HiPS structure is also observed in velocity observations (Fig.2). We have used data provided by the $\bf G$lobal $\bf O$scillation at $\bf L$ow $\bf F$requency instrument on board the SOHO satellite \cite{gab95}. It works by measuring the Doppler shifts of the NaD1/D2 line at 5889A  and 5896A. GOLF is working in single-wing  configuration, that has been changing over 12 years of observations as follows: 1)1996-1998 in the blue-wing configuration; 2)1998-2002 in the red-wing; 3) up today in the blue-wing. The nature of the measured one-wing signal remains nearly pure velocity \cite{pal}. The loss of two-wing measurements on the sodium profile induced to develop new methods to extract the Doppler velocity from the raw photometric signals \cite{gar05,ulr}. Due to the formation heights of the spectral lines GOLF observes higher up in the atmosphere compared to VIRGO \cite{jimrey}. This implies that we are sampling two different regions of the solar atmosphere in which the mode amplitudes are different, getting larger with increasing height \cite{sim,sim08a}. Fig.2 shows the high frequency part of the power spectrum from GOLF blue and red-wing configuration, where the color with which the GOLF data are shown is indicative of the observing wing. Above 5.7 mHz we found the HiPS pattern that clearly appears in the blue-wing configuration up to $\approx$7mHz. In the red-wing configuration around $\approx$6.5mHz we can spot the presence of few peaks. The possible explanation of this different behavior has to be found in the less counting rate in the red-wing configuration compared to the blue one. As a result the noise level is higher in the red-wing configuration as Fig.2 shows. In fact, in the blue-wing data from 5mHz up to 7.5mHz we have a decreasing trend of HiPS amplitude with frequency and then it stops: this is the level of noise in the blue data. In the red-wing data, instead, the decreasing trend stops at around $\approx$6mHz, thus the level of noise is higher compared to the blue one. 
\begin{figure}
\includegraphics[width=6cm,clip]{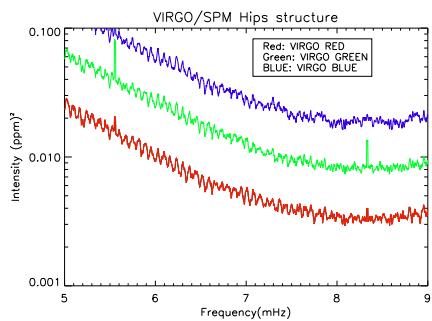}
\caption{HiPS structure from VIRGO Red, Green and Blue observations}
\label{labelfig1}
\end{figure}
\begin{figure}
\includegraphics[width=6cm,clip]{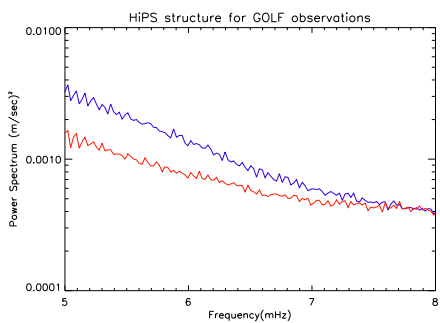}
\label{labelfig2}
\caption{HiPS structure from GOLF Blue/Red-wing configuration.}
\end{figure}
\subsection{Averaged Cross Spectrum}
To decrease the level of noise, we analyze GOLF data with the cross-spectrum technique (CS) to extract the common signal with a better signal-to-noise ratio. The AVERAGE cross spectrum is defined as the product of  the Fourier transform of one channel multiplied by the complex  conjugate of the Fourier transform of the second one and then average all the sub-series. It enhances any coherent signals between the two data sets \cite{gar99}. We can apply the cross-spectrum technique to GOLF data, because the instrument has two photo-multipliers simultaneously measuring in both configurations. Therefore, we can determine the cross spectrum by using two time series. Fig.3 shows the cross-spectrum for the blue and red-wing configurations:  the level of noise has decreased from the bottom by a factor of 10 in the red-wing configuration. This is a further confirmation of the incoherent noise coming from the instrument due to the low red-wing counting rate. This will improve the determination of HiPS amplitude variation over the solar cycle.
\begin{figure}
  \includegraphics[width=6cm,clip]{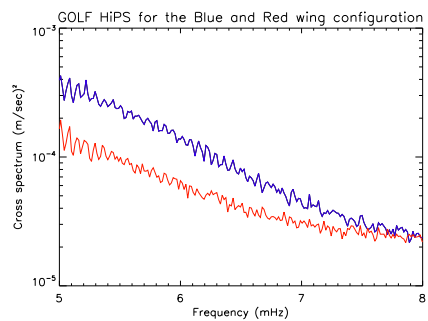}
  \caption{Yearly averaged-cross spectrum}
\end{figure}
\section{HiPS amplitude variation over solar cycle 23}
We used 9 years of GOLF and VIRGO data starting from the 25th of June 1996 and ending the 25th June 2005. We split the whole time series in subset length of 4 days to observe the HiPS. We performed the averaged cross-spectrum analysis by taking two time series in the blue and red-wing configuration of GOLF observations. Then we performed the CS in each of the individual subset and we averaged the CS over one year. With this yearly averaged CS we select a short frequency band from 5.7 up to 6.3 mHz and then we integrated the amplitude in this interval for every year from 1996 to 2006. Fig.4 shows the HiPS amplitude variation over solar cycle 23: because GOLF is observing at two different heights in the atmosphere, for the blue period we determine the amplitude variation with respect to 1996 and for the red-wing configuration with respect to 1998. In the red-wing configuration we found $\approx$18$\%$ of increasing acoustic emissivity over the period 1998-2002, while for the blue-wing period we found an increase from 1996 up to 2003 of $\approx$40$\%$. We applied the same technique to VIRGO Red and Green data to look for the presence of a significant enhancement over the ascending phase of solar cycle 23. Fig.5 shows the HiPS amplitude variation with solar cycle  as seen by the cross correlation analysis between the Green and Red channels. We choose these two channels because they are the ones that look higher up in the atmosphere compared to the blue one. As we can see within the errors we find a slight indication of an increasing acoustic emissivity, but the size of the enhancement is pretty small. Several different mechanisms have been proposed in literature in order to explain the observed enhancement above 5mHz in velocity amplitude around sunspots and active regions \cite{jai02}, but presently we believe that mode conversion from fast acoustic to fast magnetoacoustic is the preferred one in the{\it Quiet Sun} \cite{sch}. In integrated sunlight observations, even if in a sunspot the magnetic flux is many orders of magnitude larger than in a point of the quiet Sun, the integration over the solar surface will be dominated by the quiet Sun \citep{alm04}. Recently observations in the Quiet Sun show magnetic field strengths going up to 2 KG \cite{alm04}. Under this condition mode conversion can occur and the two restoring force acting to generate the fast mode (gas pressure and magnetic pressure fluctuations) are in phase and add constructively for the fast mode, increasing as a result the acoustic emissivity. Furthermore, the observed difference in intensity and velocity observations might be explained in terms of observational heights. In order for mode conversion to occur, it is of vital importance for the wave to pass through a layer where Alfven velocity equals the sound speed. For a 1KG magnetic field strength this layer is at around 60Km \cite{sch} and therefore mode conversion has not yet been occurred at the height where intensity observations are performed.  
\begin{figure}
\includegraphics[width=6cm,clip]{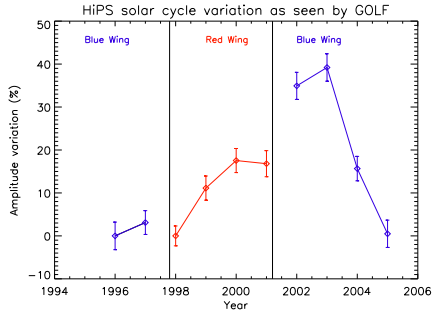}
\label{labelfig3}
\caption{HiPS amplitude variation from GOLF red and blue-wing configuration.}
\end{figure}
\begin{figure}
\includegraphics[width=6cm,clip]{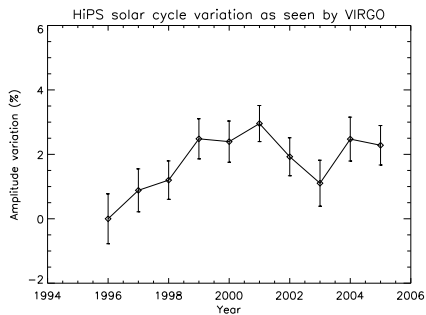}
\caption{HiPS amplitude variation from VIRGO red/green channels}
\label{labelfig4}
\end{figure}
\section{Conclusion}
We have used long high-quality integrated sunlight measurements to investigate the frequency dependence of $p$-mode surface velocity amplitudes with the solar cycle. We have shown that between 5.7<$\nu$<6.3 mHz in velocity observations there is a significant enhancement over the ascending phase of solar cycle 23.  But it is important to underline that these are preliminary results, because we are still working on the best calibration of the third GOLF velocity segment and the absolute value of the power could be slightly changed. 
 The analysis carried out in this study shows that integrated sunlight measurements have the capability to investigate the solar-cycle changes induced at high frequency. Even more these observations have shown that the magnetism of the {\it Quiet Sun} changes the nature of the acoustic waves. This needs to be taken into account, if we want to understand the role played by acoustic and/or magnetoacoustic waves in the chromospheric heating.  
This investigation has a further application in asteroseismology. The technique developed here can be used to fulfill our understanding on stellar activity \citep{kar07}. Recent mission such as Kepler aim to track variable stars for several years and this will give us the chance to follow their magnetic activity cycle by using intensity observations at low and high frequency. The analysis of surface velocity amplitude behavior over the stellar activity cycle will give us a measure of the strength of the star activity compared to the Sun.
\begin{theacknowledgments}
  This work has been partially founded by the Swiss National Funding 200020-120114, by the Spanish grant  
PENAyA2007-62650 and the CNES/GOLF grant at the Sap-CEA/Saclay. SOHO  
is an international cooperation between ESA and NASA.
\end{theacknowledgments}






\end{document}